\documentclass{PoS}

\def\beq{\begin{equation}}
\def\eeq{\end{equation}}
\def\bea{\begin{eqnarray}}
\def\eea{\end{eqnarray}}
\def\d{\displaystyle}

\def\Eq#1{Eq.~(\ref{#1})}

\title{Axigluon signatures at hadron colliders}

\ShortTitle{Axigluon signatures at hadron colliders}

\author{Germ\'an Rodrigo
\thanks{Supported by Consejo Superior de Investigaciones 
Cient\'{\i}ficas (CSIC) under grant PIE 200650I247, 
Ministerio de Educaci\'on y Ciencia (MEC) under grant FPA2007-60323
and CSD2007-00042, Generalitat Valenciana under grant GVACOMP2007-156,
and European Commission MRTN FLAVIAnet under contract MRTN-CT-2006-035482.
}\\
        Instituto de F\'{\i}sica Corpuscular,
        CSIC-Universitat de Val\`encia, \\ 
        Apartado de Correos 22085,
        E-46071 Valencia, Spain. \\
        E-mail: \email{german.rodrigo@ific.uv.es}}

\abstract{Axigluons are colored heavy neutral gauge boson that 
are predicted by some theories. The most important model-independent 
manifestation of axigluons is the generation
of a forward--backward asymmetry in top-antitop quark production
at $p\bar{p}$ collisions which originates from the charge asymmetry.
We update our previous analysis for the inclusive QCD induced
forward--backward asymmetry and define a new observable which
is more sensitive to the effect than the forward--backward asymmetry.
We find a lower limit of $1.4$~TeV at 90\% C.L. on the axigluon mass 
from recent measurements of the asymmetry at Tevatron, and 
extend the analysis to LHC in suitably selected samples. 
Like at Tevatron, the charge asymmetry can probe larger values
of the axigluon mass than the dijet mass distribution.}

\FullConference{8th International Symposium on Radiative Corrections \\
                 October 1-5, 2007\\
                 Florence, Italy}

\begin{document}

\section{Introduction}

The Large Hadron Collider (LHC) will enter into operation very soon,
allowing to explore the existence of new physics at the TeV energy scale with
unprecedented huge statistics~\cite{Gianotti:2005fm}.
Since the top quark is the heaviest known
elementary particle it plays a fundamental role in many extensions of the
Standard Model (SM), and its production and decay channels are
promising probes of new physics. The total cross section of top-antitop
quark production at LHC is about $100$ times larger than at Tevatron.
This will lead to the production of millions of $t\bar{t}$ pairs per year
even at the initial low luminosity of ${\cal L}=10^{33}$cm$^{-2}$s$^{-1}$
(equivalent to $10$~fb$^{-1}$/year integrated luminosity).

Some properties of the top quark can be studied at Tevatron
through the forward--backward asymmetry which originates from
the charge asymmetry~\cite{mynlo,Halzen:1987xd}.
The Born processes relevant for top quark
production, $q\bar{q} \to t\bar{t}$ and $gg \to t\bar{t}$, do not
discriminate between final quark and antiquark, thus predicting
identical differential distributions also for the hadronic
production process. At order $\alpha_s^3$ however a charge
asymmetry is generated and the differential distributions
of top quarks and antiquarks are no longer equal.
A similar effect leads also to a strange-antistrange
quark asymmetry, $s(x)\neq \bar{s}(x)$, through next-to-next-to-leading
(NNLO) evolution of parton densities~\cite{Catani:2004nc}.
The inclusive charge asymmetry has its origin in two different reactions:
radiative corrections to quark-antiquark annihilation (Fig.~\ref{fig:qqbar})
and interference between different amplitudes contributing
to gluon-quark scattering
$qg \to t \bar{t}q$ and $\bar{q}g \to t \bar{t}\bar{q}$.
Gluon-gluon fusion remains obviously symmetric. The integrated
forward--backward asymmetry has been predicted
to be about $+5\%$ at Tevatron~\cite{mynlo}; with the top quarks
emitted preferentially in the direction of the incoming protons.
This prediction suffers, however, from a sizable uncertainty because,
although arising from a one-loop calculation and the corresponding real
emission terms, it is still a leading order (LO) result.

At LHC the total forward--backward asymmetry vanishes trivially 
because the proton-proton initial state is symmetric.
A charge asymmetry is, however, still visible in
suitably defined distributions~\cite{mynlo}.

The forward--backward asymmetry of top quarks has already been measured
at Tevatron~\cite{Schwarz:2006ud,newcdf,Weinelt:2006mh,d0}.
The latest CDF analysis~\cite{newcdf} based on $1.7$~fb$^{-1}$ 
integrated luminosity, which supersedes the results presented 
in Ref.~\cite{Weinelt:2006mh}, gives for the inclusive asymmetry
\beq
A(\Delta y\cdot Q_l) = 0.28 \pm 0.13~(\rm{stat}) \pm 0.05~(\rm{sys})~,
\label{eq:newcdf}
\eeq
where the charge asymmetry is defined by the difference in the
number of events with positive and negative $\Delta y\cdot Q_l$,
the rapidity difference of the semileptonically and hadronically
decaying top quark times the charge of the charged lepton.
The inclusive asymmetry,
although compatible with the theoretical prediction, is still
statistically dominated. The statistical error,
is expected to be reduced to $0.04$ with
$8$~fb$^{-1}$~\cite{Schwarz:2006ud},
which is comparable with the systematic error.

The measurement at D0~\cite{d0} with $0.9$~fb$^{-1}$
integrated luminosity gives for the uncorrected asymmetry
\beq
A^{\rm{obs}}_{\rm FB} = 0.12 \pm 0.08~(\rm{stat}) \pm 0.01~(\rm{sys})~.
\eeq
Like CDF, this analysis use $y_t-y_{\bar t}$ as sensitive variable.
In Ref.~\cite{d0} upper limits on $t\bar t+X$ production via a $Z'$ 
resonance are also provided. 
Measurements of the exclusive asymmetry of the four- and
five-jet samples are also given in both analysis.

Models which extend the standard color gauge group
to $SU(3)_L \times SU(3)_R$ at high energies,
the so called chiral color theories~\cite{chiralcolor},
predict the existence of a massive, color-octet
gauge boson, the axigluon, which couples to quarks
with an axial vector structure and the same strong
interaction coupling strength as QCD.
Although there are many different implementations of
chiral color theories with new particles in varying representations
of the gauge groups, the most important model-independent
prediction of these models is the existence of the axigluon,
where its main signature is the appearance of a
charge asymmetry of order $\alpha_s^2$.
Because the coupling of the axigluon to quarks is an axial
vector coupling the charge asymmetry that can be generated
is maximal.

\section{The QCD induced charge asymmetry}

\begin{figure}[th]
\begin{center}
\includegraphics[width=7cm]{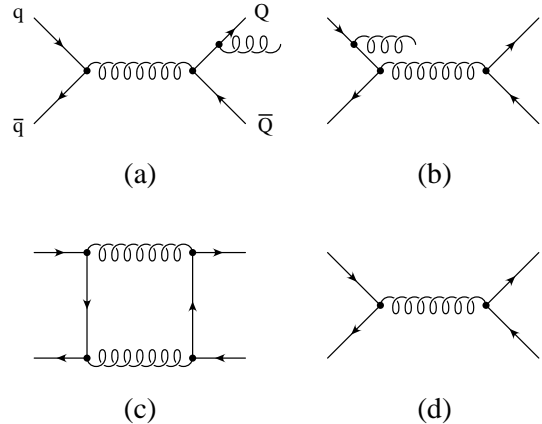}
\caption{Origin of the QCD charge asymmetry in hadroproduction
of heavy quarks: interference of final-state
(a) with initial-state (b) gluon bremsstrahlung,
plus interference of the double virtual gluon exchange (c)
with the Born diagram (d). Only representative diagrams are shown.}
\label{fig:qqbar}
\end{center}
\end{figure}

The QCD induced charge asymmetry in the reaction $q\bar{q} \to t\bar{t} (g)$
is generated by the interference of final-state with initial-state
gluon radiation  [Fig.~\ref{fig:qqbar}, (a)$\times$(b)] and
by the interference of virtual box diagrams with
the Born process [Fig.~\ref{fig:qqbar}, (c)$\times$(d)].
The virtual plus soft radiation on one hand and the real
hard radiation on the other contribute with opposite signs,
with the former always larger that the latter such that
the inclusive asymmetry becomes positive. Top quarks are
thus preferentially emitted in the direction of the
incoming quark at the partonic level, which translates to
a preference in the direction of the incoming proton
in $p\bar{p}$ collisions.
Flavour excitation $gq(\bar{q}) \to t\bar{t}X$ generates
already at tree-level a forward--backward asymmetry which at Tevatron
is also positive although one order of magnitude smaller than
the asymmetry from $q\bar{q}$ annihilation.

\begin{figure}[th]
\begin{center}
\includegraphics[width=7cm]{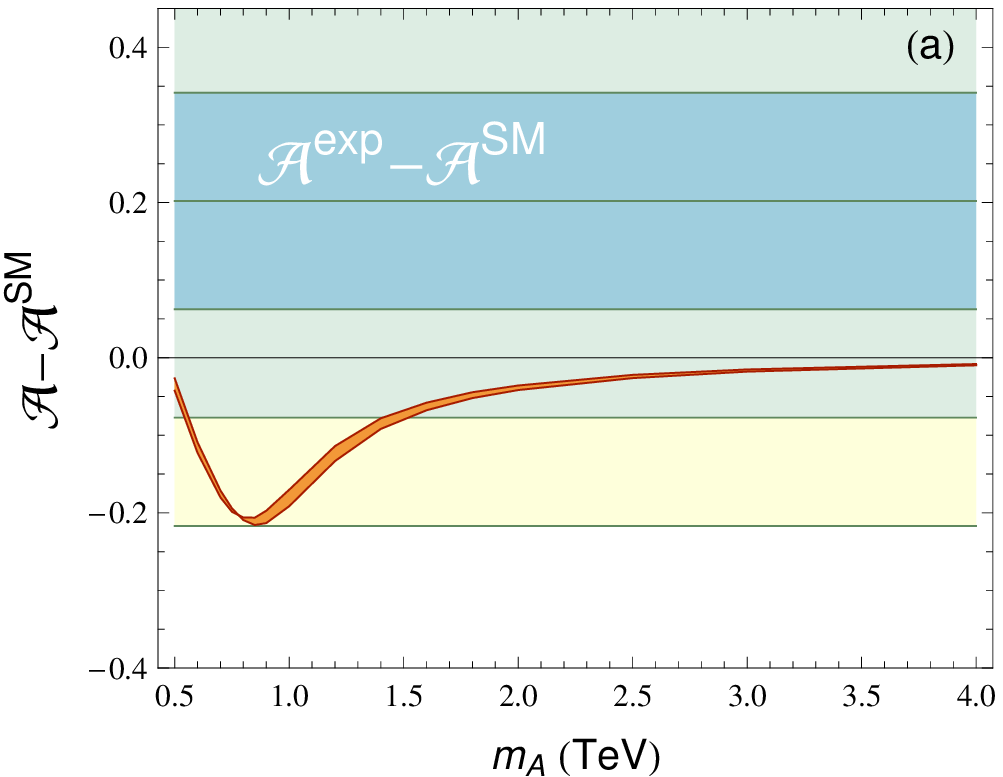}
\includegraphics[width=7cm]{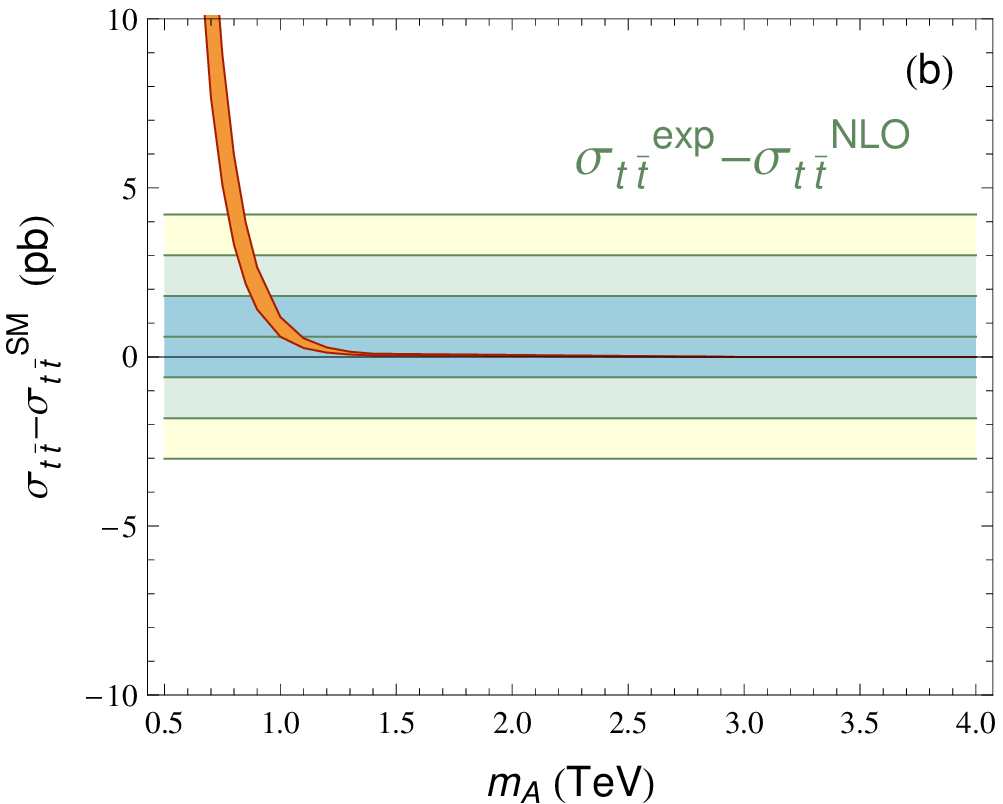}
\caption{Comparison of the axigluon contribution 
to the top quark pair asymmetry (a), and to 
the top-antitop total cross section (b) with the $1\sigma$, 
$2\sigma$ and  $3\sigma$ contours as a function of the axigluon mass.}
\label{fig:axi}
\end{center}
\end{figure}

Updating our previous analysis~\cite{mynlo}, we predict 
for the total charge asymmetry at $\sqrt{s}=1.96$~TeV: 
\beq
A =\frac{N_t(y\ge 0)-N_{\bar{t}}(y\ge 0)}
{N_t(y\ge 0)+N_{\bar{t}}(y\ge 0)} = 0.051(6)~.
\label{ourprediction}
\eeq
We have also proposed~\cite{Antunano:2007da} 
a new differential distribution, the pair asymmetry, 
that leads to an enhancement of the charge asymmetry.
Selecting events where the rapidities 
$y_+$ and $y_-$ of both the top and antitop quarks
have been determined, we define $Y=1/2(y_+ + y_-)$ 
as average rapidity, and consider the differential 
pair asymmetry ${\cal A}(Y)$ for all events with fixed $Y$
as a function of $Y$. 
For the integrated pair asymmetry we predict~\cite{Antunano:2007da}: 
\beq
{\cal A} = 
\frac{\d \int dY (N_\mathrm{ev.}(y_+>y_-) - N_\mathrm{ev.}(y_+<y_-))}
{\d \int dY (N_\mathrm{ev.}(y_+>y_-) + N_\mathrm{ev.}(y_+<y_-))}
=  0.078(9)~.
\label{eq:pair}
\eeq
The integrated pair asymmetry is equivalent to the definition 
of the asymmetry used in Refs.~\cite{newcdf,d0}. 
The reason for the enhancement of the effect can be understood 
as follows: by defining the pair
asymmetry one essentially investigates the forward--backward 
asymmetry in the $t\bar{t}$ rest frame, where the forward--backward 
asymmetry amounts to $7-8.5$\%~\cite{mynlo}, depending on $\hat{s}$. 
This value is largely recovered by considering the pair 
asymmetry ${\cal A}(Y)$, independently of $Y$. In contrast, events 
where both $t$ and $\bar{t}$ are produced with positive and 
negative rapidities do not contribute to the integrated 
forward--backward asymmetry $A$, which is therefore reduced 
to around $5$\%.

\section{Axigluon limits from Tevatron}

The interference between the gluon and axigluon induced amplitudes
for the reaction $q\bar{q} \to t\bar{t}$ does not 
contribute to the production cross section. However, it generates a charge 
asymmetry that gives rise to a forward--backward asymmetry 
in $p\bar{p}$ collisions in the laboratory frame~\cite{Sehgal:1987wi}. 
The square of the axigluon amplitude is symmetric and contributes 
to the total cross-section, which will show a typical resonance 
peak in the top-antitop invariant 
mass distribution~\cite{Bagger:1987fz}. 
While the interference term is suppressed by 
the squared axigluon mass $1/m_A^2$, the contribution of the 
square of the axigluon amplitude will be suppressed by $1/m_A^4$. 
It is therefore obvious that the forward--backward asymmetry is 
potentially sensitive to larger values of the axigluon mass
than the top-antitop dijet distribution. 
Gluon-gluon fusion is not affected by the axigluon exchange 
because there are no direct gluon-axigluon vertices with an odd 
number of axigluons~\cite{Bagger:1987fz} due to parity. 

\begin{figure}[ht]
\begin{center}
\includegraphics[width=6.7cm]{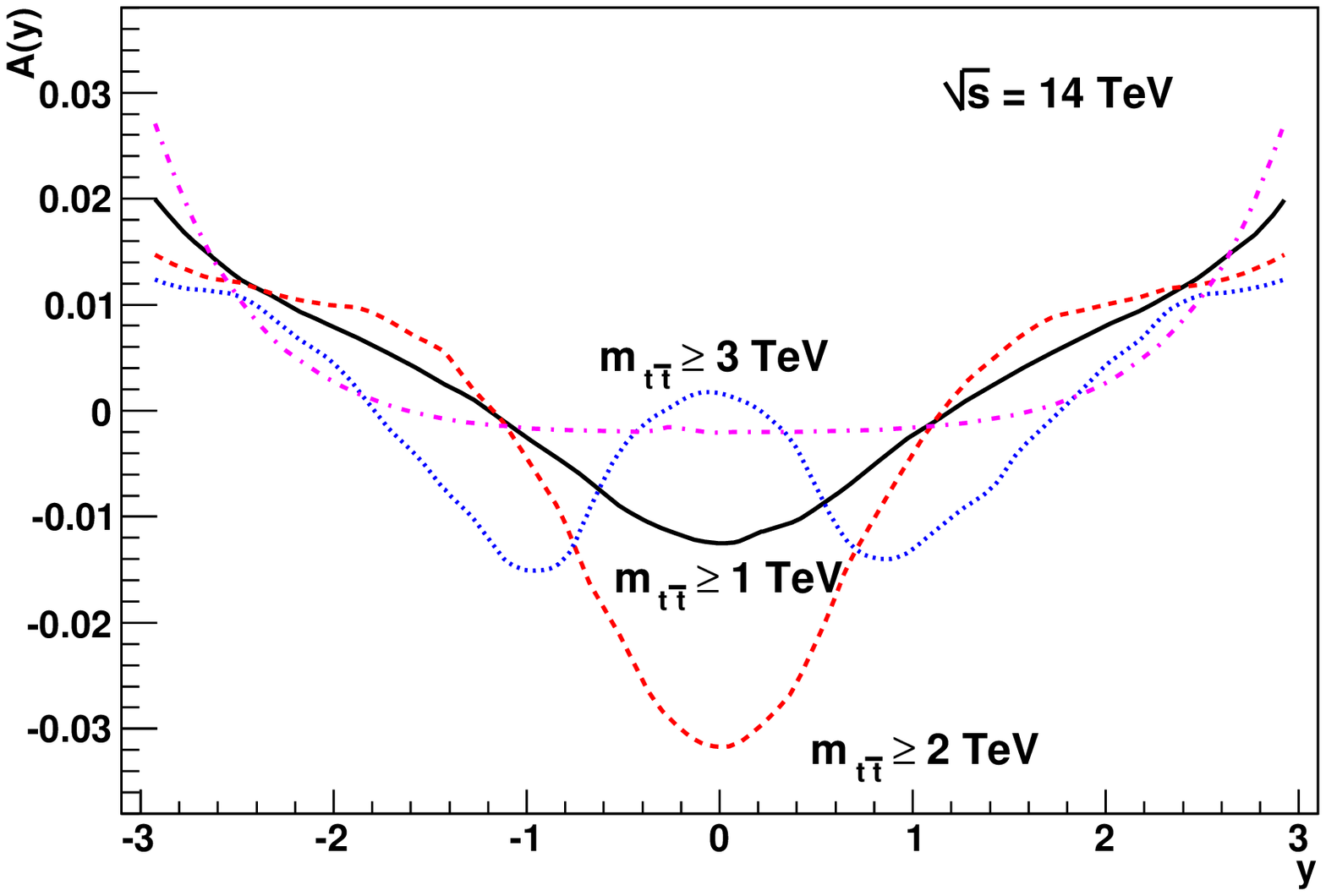}
\includegraphics[width=6.6cm]{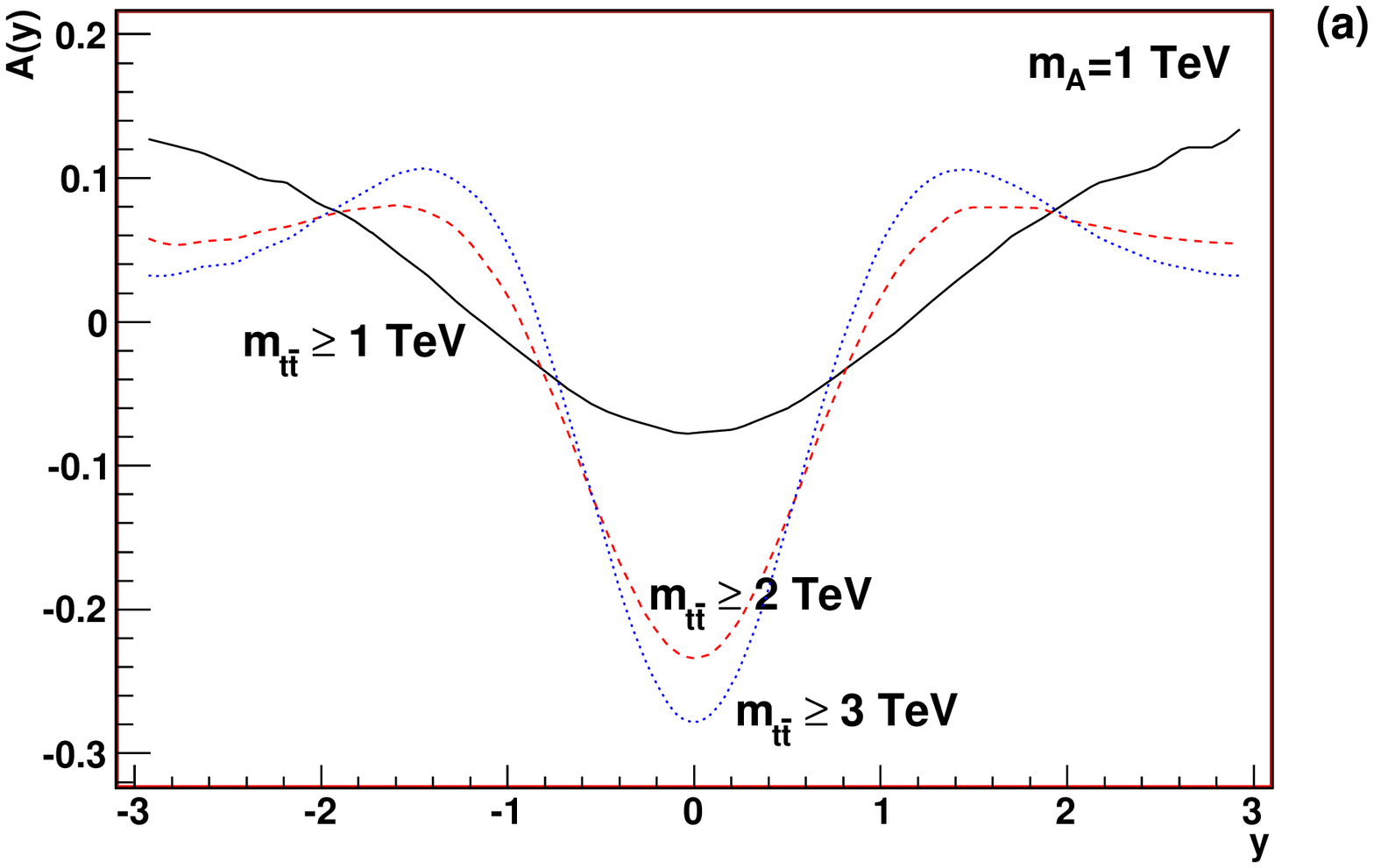}
\centering
\caption{\label{fig:lhcQCDasym} Differential top quark charge 
asymmetry at LHC generated by QCD (left) and by the exchange of 
a massive axigluon (right) for top-antitop quark invariant masses 
larger than $2m_t$ (dotted-dashed), $1$~TeV (solid), $2$~TeV (dashed), 
and $3$~TeV (dotted). 
Factorization and renormalization scales set to $\mu=m_t$.}
\end{center}
\end{figure}

Figs.~\ref{fig:axi}(a) and~\ref{fig:axi}(b) show  
the contribution of axigluon exchange in comparison to the experimental 
measurement of the integrated pair asymmetry (\Eq{eq:newcdf}), 
and of the top-antitop total cross section~\cite{topcross},
$\sigma_{t\bar t} = 7.3 \pm 0.9$~(pb), respectively. 
For the SM prediction we use \Eq{eq:pair}, and 
$\sigma_{t\bar t}^{\rm{NLO}} = 6.7 \pm 0.8$~(pb)~\cite{Cacciari:2003fi}.

We observe that above $m_A>1$~TeV the contribution 
of the axigluon to the total cross section is almost 
suppressed. Better determinations of the total cross 
section will not lead to a significant improvement 
in the bound of the axigluon mass. On the 
contrary, the pair asymmetry is particularly sensitive to 
axigluon masses below 2-2.5 TeV, and little improvements 
can lead to a significant change in the lower bound. 
We can establish a lower bound of $m_A>1.4$~TeV at 
90\% C.L.

\section{QCD and axigluon induced asymmetries at the LHC}

Top quark production at LHC is forward--backward symmetric in the 
laboratory frame as a consequence of the symmetric colliding 
proton-proton initial state. Furthermore, the total cross section 
is dominated by gluon-gluon fusion and thus the charge asymmetry 
generated from the $q\bar{q}$ and $gq$ ($g\bar{q}$) reactions is
negligible in most of the kinematic phase-space. The effect 
can be studied nevertheless by selecting appropriately chosen 
kinematic regions.  
We have therefore analyzed the effect of selecting samples with 
high invariant masses of the top-antitop quark pair. Those samples 
should have a higher amount of $q\bar{q}$ induced events, 
and an enhanced axigluon contribution even for large $m_A$.  
Thus a sizable asymmetry is expected, although at the 
price of reducing the total event rate. This should not 
be a problem at LHC due to the huge top-antitop quark 
yields. 

Fig.~\ref{fig:lhcQCDasym} illustrates the QCD and the axigluon 
contributions to the differential charge asymmetry as a function of 
the top quark rapidity. We have defined a new charge asymmetry 
where only the the central region is taken into account:
\beq
A_C(y_C) = \frac{\sigma_t(|y|\le y_C)-\sigma_{\bar{t}}(|y|\le y_C)}
{\sigma_t(|y|\le y_C)+\sigma_{\bar{t}}(|y|\le y_C)}~.
\eeq
Notice that $A_C(y_C)$ vanishes if the whole rapidity spectrum
is integrated. A maximum is reached at about $y_C=1$. 
Predictions for different kinematical cuts 
and axigluon masses can be found in Ref.~\cite{Antunano:2007da}.

\section{Summary}

We have updated our previous analysis of the forward--backward and
the charge asymmetry in top quark production at hadron colliders.
We have also proposed a new observable, the pair asymmetry,
where the effect at the Tevatron is enhanced by about a factor~$1.5$.
We have shown that the pair asymmetry is more sensitive to larger 
axigluon masses that the total top-antitop quark cross section. 
Preliminary results on the charge asymmetry from the
Tevatron lead to a limit on the axigluon mass of~$1.4$~TeV
at 90\% C.L.
The analysis has been extended to the LHC. Restricting the event
sample to regions of large $t\bar{t}$ mass 
large axigluon masses can be explored.

\section*{Acknowledgements}

I want to thank J.H. K\"uhn and O. Antu\~nano for a fruitful 
collaboration, and the local organizing Committee of RADCOR07 
and the Galileo Galilei Institute for Theoretical Physics
for their kind hospitality.

\end{document}